\documentclass[11pt,a4paper]{scrartcl}
\pdfoutput=1

\usepackage{CLICdp}

\usepackage{CLICdp_definitions}
\usepackage[symbol]{footmisc}
\usepackage{feynmf}
\def \couplingunc {19}
\newcommand{\wwfusion}{\ensuremath{\PW\PW} fusion\xspace}
\newcommand{\hnn}{\ensuremath{\PH\PGn\PGn}\xspace}
\newcommand{\hmumu}{\ensuremath{\PH \to \mpmm}\xspace}
\newcommand{\eemm}{\ensuremath{\epem \to \epem \mpmm}\xspace}
\newcommand{\egemm}{\ensuremath{\Pepm \PGg \to \Pepm \mpmm}\xspace}
\newcommand{\nnmm}{\ensuremath{\epem \to \nuenuebar \mpmm}\xspace}
\newcommand{\ggnnmm}{\ensuremath{\gamgam \to \PGnGm\PAGnGm \mpmm}\xspace}
\newcommand{\egennmm}{\ensuremath{\Pepm \PGg \to \Pepm \PGnGm\PAGnGm \mpmm}\xspace}
\newcommand{\eennmm}{\ensuremath{\epem \to \epem \PGnGm\PAGnGm \mpmm}\xspace}
\newcommand{\BR}{\ensuremath{\text{BR}}}
\newcommand{\bs}{Beamstrahlung\xspace}
\newcommand{\sigmaBR}{\ensuremath{\sigma(\hnn) \times \BR(\hmumu)}\xspace}
\newcommand{\GH}{\ensuremath{\Gamma_{\PH}}\xspace}
\newcommand{\ud}{\,\mathrm{d}}

\ifcsname'percent'\endcsname
  \renewcommand{\percent}{\ensuremath{\,\%}\xspace}
\else 
  \newcommand{\percent}{\ensuremath{\,\%}\xspace}
\fi

\newacro{MVA}[MVA]{Multivariate Analysis}
\newacro{ISR}[ISR]{Initial State Radiation}
\newacro{TMVA}[TMVA]{Toolkit for the Multivariate Analysis}
\newacro{CM}[CM]{centre-of-mass}
\newacro{MC}[MC]{Monte-Carlo}
\newacro{EPA}[EPA]{Effective Photon Approximation}
\newacro{BDT}[BDT]{Boosted Decision Tree}
\newacro{SM}[SM]{Standard Model}
\newacro{PDF}[PDF]{Probability Density Function}
\newacro{CLIC}[CLIC]{Compact Linear Collider}

\title{Physics potential for the measurement of $\sigma(\hnn) \times \BR(\hmumu)$ at a 1.4 TeV CLIC collider}

\clicdpnote{2014}{005}  
\date{\today}

\addauthor{G.~Milutinović-Dumbelović}{\institute{1}}
\addauthor{I.~Božović-Jelisavčić}{\institute{1}}
\addauthor{C.~Grefe}{\institute{2}}
\addauthor{S.~Lukić\thanks{slukic@vinca.rs}}{\institute{1}}
\addauthor{M.~Pandurović}{\institute{1}}
\addauthor{P.~Roloff}{\institute{2}}

\addinstitute{1}{Vinča Institute, University of Belgrade, Serbia}
\addinstitute{2}{CERN, Switzerland}

\abstract{The potential for the measurement of the branching ratio of the Standard Model-like Higgs boson decay into a \mpmm pair at 1.4 TeV CLIC is analysed. The study is performed using the fully simulated \clicild detector concept, taking into consideration all the relevant physics and the beam-induced backgrounds. Despite the very low branching ratio of the \hmumu decay, we show that the product of the branching ratio times the Higgs production cross section can be measured with a statistical uncertainty of 38\percent, assuming an integrated luminosity of 1.5~\abinv collected in five years of the detector operation at the 1.4 TeV CLIC with unpolarised beams. With polarised beams (+80\percent, -30\percent), the statistical uncertainty is better than 25\percent}

\titlecomment{This work was carried out in the framework of the CLICdp collaboration}

\graphicspath{ {./logos/}{./figures/} }

\begin{document}

\titlepage

\section{Introduction}
Measurements of Higgs branching ratios, and consequently Higgs couplings, provide a strong test of the \ac{SM} and the physics beyond. Models that could possibly extend \ac{SM} Higgs sector, such as the Two Higgs Doublet model, the Little Higgs model or the Compositeness models predict Higgs couplings to EW bosons and Higgs Yukawa couplings (coupling-mass linearity) that deviate from the \ac{SM} predictions \cite{Gupta12, Englert14}. 

The \ac{CLIC} represents an excellent environment to study properties of the Higgs boson, including the couplings, with a very high precision \cite{CLIC_PhysDet_CDR, CLIC_snowmass13}. Measurement of the rare \hmumu decay is particularly challenging because of the very low branching ratio (\BR) of $2.14 \times 10^{-4}$ predicted in the \ac{SM} at the Higgs mass of 126 GeV \cite{Dittmaier:2012vm}. Presently the search for the \hmumu decay at \mbox{ATLAS} including 25~\fbinv of data has yielded an upper limit for \BR(\hmumu) of $1.5 \times 10^{-3}$ \cite{ATLAS-hmumu-2014}. A similar upper limit was reported by the CMS experiment as well \cite{CMS-hmumu-2014}. In the projections for the HL-LHC \mbox{ATLAS} experiment, the \ac{SM} expectations are $2.3 \sigma$ signal significance, or 46\percent signal counting uncertainty with $300\,\fbinv$ of data, and $7\sigma$ signal significance, or 21\percent counting uncertainty with $3\,\abinv$ of collected data \cite{ATLAS-HL-LHC-Higgs}. The respective projected significances at CMS are $2.5 \sigma$ with $300\,\fbinv$ of data, and $7.9\sigma$ with $3\,\abinv$ \cite{CMS-hmumu-2014}.
The measurement at \ac{CLIC} requires excellent muon identification efficiency and momentum resolution as well as efficient background rejection.

One of the possible staged scenarios of the \ac{CLIC} construction and operation, optimized for the best physics reach in shortest time with optimal cost, comprises the $350\unit{GeV}$, $1.4\unit{TeV}$ and $3\unit{TeV}$ \ac{CM} energy stages \cite{CLICCDR_vol3}. In the latter two stages sufficiently large Higgs boson samples are produced via the \wwfusion process that the measurement of the rare \hmumu decay can be performed. The analysis of the \hmumu decay at $3\unit{TeV}$ is presented in Ref. \cite{lcd:grefeHmumu2011}. The analysis at $1.4\unit{TeV}$ is presented here. The target value of the integrated luminosity at the 1.4\unit{TeV} stage is $1.5\,\abinv$. At the instantaneous luminosity of $3.2\times10^{34} \unit{cm}^{-2} \unit{s}^{-1}$ this is achieved in approximately five years of physics operation with 200 running days per year and an effective up-time of 50\percent. This analysis assumes that the total integrated luminosity is collected with the \clicild detector concept \cite{LCDnote_CLICILDCDRgeo}. Unpolarised beams are assumed. With 80\percent left-handed beam polarisation for the electrons and 30\percent right handed polarisation for the positrons, the Higgs production through \wwfusion can be enhanced by a factor 2.34 \cite{CLIC_snowmass13}. A conservative estimate of the final uncertainty in the case that $1.5\,\abinv$ of data is collected with polarized beams is given at the end.

In this note, the analysis procedure is presented, and statistical and systematic uncertainties of the measurement are determined and discussed. In Section \ref{sec-simulation} the simulation tools used for the analysis are listed and in Section \ref{sec-detector} the CLIC\_ILD detector model is briefly described. Signal and background processes and event samples are discussed in Section \ref{sec-samples}. Tagging of spectator electrons in background processes is described in Section \ref{sec-tagging}. Event preselection, and a final selection using \ac{MVA} is described in Section \ref{sec-selection}. The di-muon invariant mass fit and the extraction of the statistical uncertainty of the measurement are described in Section \ref{sec-fit}. A brief discussion of the eventual benefit of a better transverse momentum, \pT, resolution is found in Section \ref{sec-pTres}. Systematic uncertainties are discussed in Section \ref{sec-systematics}, followed by the conclusions. In the Appendix, distributions of sensitive variables used in the \ac{MVA} are shown for all processes before and after the selection.

\section{Simulation and analysis tools}
\label{sec-simulation}
The Higgs production through \wwfusion was simulated in \whizard V 1.95 \cite{Kilian:2007gr} including the CLIC beam spectrum and the \ac{ISR}. \pythia 6.4 \cite{Sjostrand2006} was used to simulate the Higgs decay into two muons. The background events were also generated in \whizard, using \pythia 6.4 to simulate the hadronisation and fragmentation processes. The CLIC luminosity spectrum and the beam induced processes were obtained from \guineapig 1.4.4 \cite{Schulte:1999tx}. 

The interactions with the detector were simulated using the CLIC\_ILD detector model within the \mokka simulation package \cite{Mora2002}, based on \geant \cite{Agostinelli2003}. A particle flow algorithm \cite{thomson:pandora, Marshall2013153} was employed in the reconstruction of the final state particles within the Marlin reconstruction framework \cite{MarlinReco}. 

The \ac{TMVA} package \cite{TMVA:2010} was used for the multivariate classification of signal and background events using their kinematic properties.

\section{The CLIC\_ILD detector model}
\label{sec-detector}
For CLIC, the ILD detector model \cite{ildloi:2009} has been modified according to specific experimental conditions at CLIC \cite{CLIC_PhysDet_CDR}. The subsystems of particular relevance for the present analysis are discussed in the following. 

The main tracking device of CLIC\_ILD is a Time Projection Chamber (TPC) providing a point resolution in the \rphi plane better than $100 \unit{\upmu m}$. Additional silicon trackers provide precision tracking at the outer surface of the TPC with single point accuracy in the \rphi direction of $7 \unit{\upmu m}$. Vertex detectors closer to the beam-pipe provide resolution better than $3 \unit{\upmu m}$ \cite{CLIC_PhysDet_CDR}.

High efficiency muon identification is made possible by the iron yoke instrumented with 9 layers of RPC detectors. Muon momenta are determined using the matching tracks in the TPC and the silicon trackers. The efficiency of the muon identification is influenced by the longitudinal segmentation of iron, as well as by the background processes. In the sample of muons from the \hmumu decays, muon efficiency is above 99\percent in the barrel region. 

This analysis depends particularly on the muon momentum resolution as it influences the width of the reconstructed di-muon invariant mass peak. The average muon transverse momentum resolution for the signal sample in the barrel region is $\Delta(1/\pT) = 3.3\times10^{-5} \unit{GeV}^{-1}$. 

In the very forward region of the CLIC\_ILD detector, below 8\degrees, no tracking information nor hadronic calorimetry is available. The region between 0.6\degrees and 6.3\degrees is instrumented with two silicon-tungsten sampling calorimeters, LumiCal and BeamCal, for the luminosity measurement, beam-parameter control, as well as tagging of high-energy electrons escaping the main detector at low angles. Together with the very forward segments of ECAL, EM calorimetry is thus available in the region between 0.6\degrees and 8\degrees, offering the possibility to suppress four-fermion \ac{SM} processes with the characteristic low-angle electron signature. The simulation of the very-forward electron tagging is described in Section \ref{sec-tagging}.

Beamstrahlung photons, emitted in beam-beam interactions produce incoherent \epem pairs deposited mainly in the low-angle calorimeters. In addition, about 1.3 interactions of \bs photons per bunch crossing produce hadrons with a wide angular distribution, influencing the muon momentum resolution due to the occupancy of the inner tracker. These hadrons were included in the analysis by overlaying a map of energy deposits in all detectors, randomly picked from a pre-simulated data set, before the digitisation phase and event reconstruction. Signal and all other background processes are fully simulated in the detector.

\section{Event samples}
\label{sec-samples}
\begin{figure}
\unitlength = 1mm
\vspace{3mm}
\centering

\input{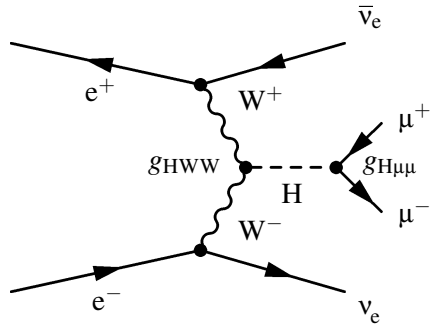}

\vspace{3mm}
\caption{\label{fig-WWfusion} Feynman diagram of the Higgs production in \wwfusion and the subsequent decay to a pair of muons.}
\end{figure}

\begin{figure}
\centering
\includegraphics[width=7cm]{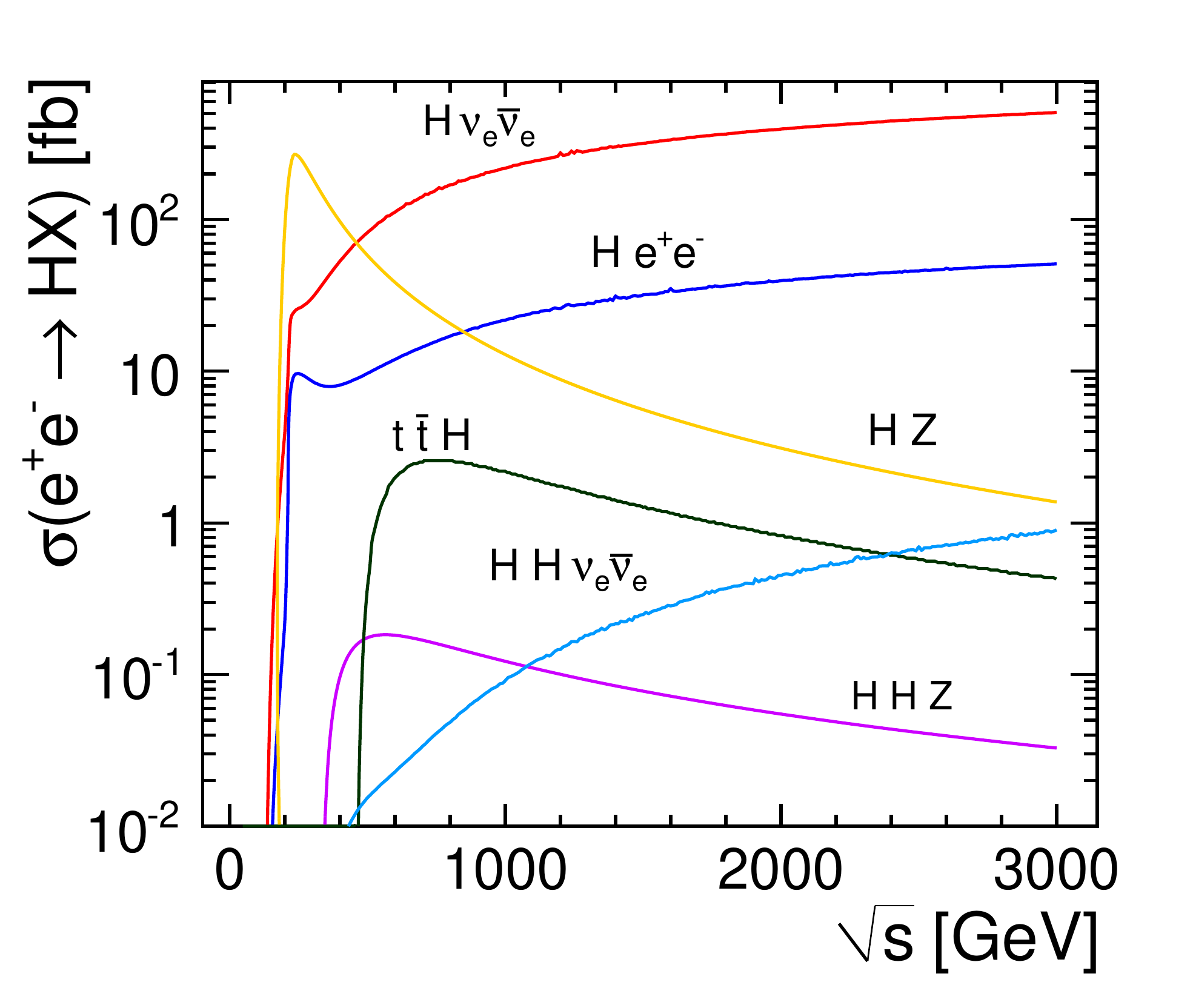}
\caption{\label{fig-xs} Higgs production cross-sections at different \ac{CM} energies.}
\end{figure}

At 1.4 TeV CLIC, the Standard Model Higgs boson is predominantly produced via \wwfusion (Figures \ref{fig-WWfusion} and \ref{fig-xs}). The effective cross-section for the Higgs production in \wwfusion assuming CLIC luminosity spectrum is 244~\unit{fb}. The Higgs production cross section above 1 TeV can be determined with a statistical precision better than 1\percent \cite{CLIC_snowmass13}. Nevertheless, the \hmumu signal statistics is expected to be small because the \ac{SM} prediction for the branching fraction of this particular decay is of the order of $10^{-4}$. 

As seen in Figure \ref{fig-xs}, at 1.4~\unit{TeV} the Higgs boson is also produced via \zz fusion, with a cross-section equal to about 10\percent of the Higgs production cross section in \wwfusion. However, on a test sample of 300 events of \zz fusion followed by the Higgs decay to a pair of muons, not a single event passed the requirements applied in this analysis (Sec. 4.3 and 4.4) implying a selection efficiency smaller than 1.2\percent (95\percent CL) for this channel. Therefore this production channel was neglected in the analysis. 

The number of simulated signal and background events was chosen so that accurate di-muon invariant mass distributions can be extracted. A sample of 24 000 signal events was simulated, roughly corresponding to a 300-fold of the number of events in 1.5~\abinv of data. From the signal sample, 6000 events were reserved for the training of the \ac{MVA}. For each of the background processes, a 2~\abinv sample was generated, of which 0.5~\abinv were reserved for the \ac{MVA} training. The full list of physics and beam-induced backgrounds is given in Table \ref{tab-processes}.

Events with momentum transfer $Q = \sqrt{\left| (p^\mu(\Pe_{out}) - p^\mu(\Pe_{in}) )^2 \right| }$\footnote{$p^\mu$ denotes the four-momentum of the particle} smaller than 4~\unit{GeV} between the incident and the spectator electron, $\Pe_{in}$ and $\Pe_{out}$, respectively, were simulated using the \ac{EPA}. In the \ac{EPA} approach, the spectator electron is substituted by a quasi-real photon. In the presentation of results of this analysis, such events will be grouped together with the processes involving \bs photons with the analogous initial state. Thus, for example, the sample denoted \egemm contains both the events in which the photon is a \bs photon, and the events in which the photon originates from the EPA modelling of the \eemm process with small momentum transfer. In this way, processes with roughly similar kinematic characteristics are grouped together. 

Table \ref{tab-processes} lists the signal and all background processes with their respective cross-sections. Cross sections for processes with spectator electrons were calculated with the cut on momentum transfer $Q > 4 \unit{GeV}$, while cross-sections for processes involving initial photons include the cross-sections for the EPA approximation of the processes with spectator electrons for $Q < 4 \unit{GeV}$. Beside that, the cross sections for processes \egemm and \egennmm represent the sum of processes with the electron and the positron in the initial state.

The process \nnmm, characterised by the same final state as the signal, as well as by the same distribution of the \ac{CM} energy in the initial state, represents an irreducible background and can not be substantially suppressed before the invariant mass fit. The process \ggnnmm has a similar final state, but a different \ac{CM} energy distribution in the initial state, since it involves \bs or \ac{EPA} photons rather than electrons. This leads to a different distribution of the boost of the di-muon system, as well as of the helicity angle, allowing separation from the signal to some extent (Section \ref{sec-mva}). 

The four-fermion production process \eemm is realised dominantly through the two-photon exchange mechanism and it fakes the missing energy signature of the signal in events in which electron spectators are emitted at angles smaller than 8\degrees. For that reason, the tagging of EM showers at low angles (see Section \ref{sec-tagging}) is applied. The processes \egemm and \egennmm are suppressed by EM shower tagging as well (see Table \ref{tab-tagging}).

\begin{table}
\centering
\begin{tabular}{| l | c |}
\hline
Process & $\sigma (\unit{fb})$  \\
\hline
$\epem \to \hnn$, \hmumu & 0.0522 \\
\nnmm & 129 \\ %
\eemm & 24.5$^*$ \\ %
\egemm & 1098$^*$ \\ %
\egennmm & 30 \\ %
\ggnnmm  & 162 \\ %
\eennmm  & 1.6 \\ %
\hline
\end{tabular}
\caption{\label{tab-processes} List of the analysed processes with the corresponding cross-sections. Cross-section values marked by * were calculated with additional kinematic requirements, $100 \unit{GeV} < m_{\mumu} < 150 \unit{GeV}$, and $8\degrees < \theta_{\PGm} < 172\degrees$ for both muons. The cross sections for all processes with photons in the initial state include both the cross sections with the \bs and with the \ac{EPA} photons. Beside that, the cross sections for processes \egemm and \egennmm represent the sum of processes with the electron and the positron in the initial state.}
\end{table}

\section{Tagging of EM showers in the very forward region}
\label{sec-tagging}
In the region below 8\degrees, the tracking information, as well as hadronic calorimetry, are not available. Background processes involving spectator electrons escaping near the beam tube mimick the missing energy signature of the signal. For that reason, as well as for the luminosity measurement and beam diagnostics, the very forward region is instrumented with EM calorimeters LumiCal and BeamCal \cite{Abramowicz:2010bg}.

\begin{figure}
\centering
\includegraphics[width=7cm]{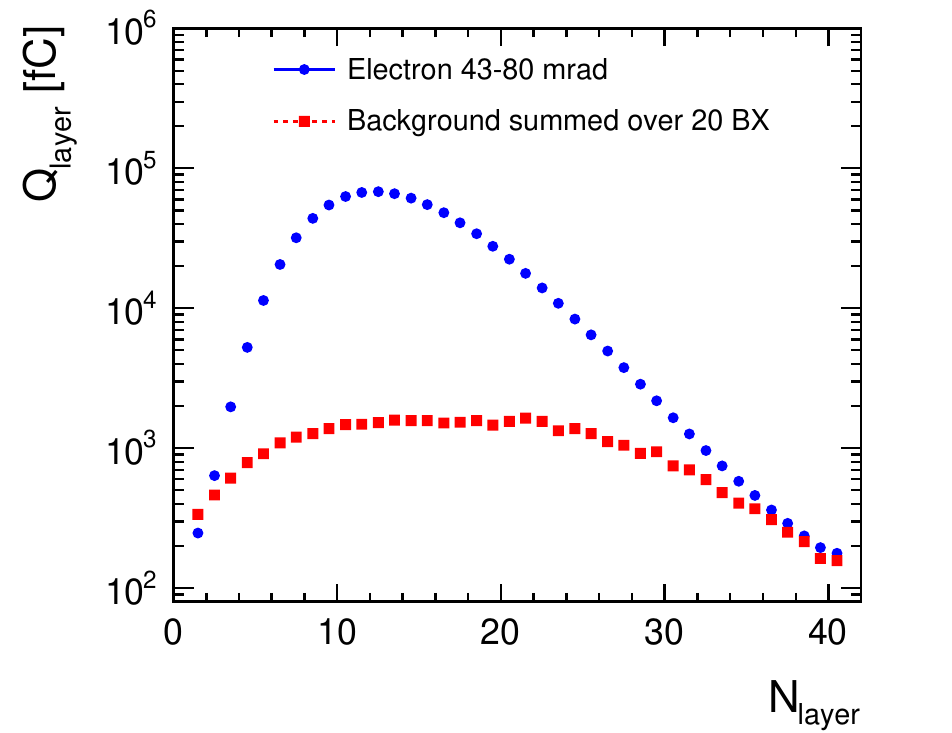}
\caption{\label{fig-layers} Comparison of the longitudinal profiles of energy deposition by EM showers and by the beam-induced background in the LumiCal at CLIC. Taken from ref. \cite{Schwartz_thesis}.}
\end{figure}

Electron detection in the very forward region involves the reconstruction of EM showers in the presence of intense beam-induced background consisting of a large number of low-energy particles, mostly incoherent pairs and hadrons \cite{dannheimsailer2011}. Neither reconstruction algorithms for the very forward detectors nor fully simulated samples of the beam-induced background were available at the time of this analysis. For that reason, tagging probability was simulated by parameterisation of the distribution of the reconstructed energy fluctuations due to the presence of the background and to the intrinsic energy resolution of the calorimeters. A requirement was imposed that the energy of the particle after adding the random fluctuation is above the mean background level in the layer with maximum deposit by at least $4\;\sigma$ of the background fluctuation. Figure \ref{fig-layers} shows the longitudinal profiles of energy deposition by EM showers and by the beam-induced background in the LumiCal at CLIC. One can readily see that the background profile is fairly flat in the region of the maximum of the signal profile. On the other hand, the width at half maximum of the signal profile is about ten layers. Thus the $4\,\sigma$ requirement above roughly corresponds to the requirement that the signal is $2\,\sigma$ above background in ten consecutive layers. 
This is an \textit{ad-hoc} but conservative estimate of the effective energy threshold due to the beam-induced backgrounds. As shown in the following, an additional ultimate energy cut at significantly higher energy was imposed to suppress tagging of signal events in coincidence with Bhabha particles.

An adverse consequence of the rejection of events with high-energy electrons in the very forward region is that individual final-state particles from Bhabha events, detected in random coincidence with other processes, cause a loss of statistics because of indiscriminate rejection of a relatively large fraction of physics events. Taking into account the boost of the Bhabha event \ac{CM} frame due to beam-beam effects, as well as the 0.5 ns bunch spacing at CLIC, and assuming a digitizer timing step of 10 ns, more than 30\percent of all events can be expected to be falsely tagged because of coincident detection of at least one final particle from a Bhabha event. In order to reduce the rate of coincident tagging of Bhabha particles, tagging was restricted to showers with energy higher than 200 GeV and polar angle above 1.7\degrees only. Under these conditions, the accidental tagging rate of Bhabha particles drops to 7\percent. Table \ref{tab-tagging} summarises the rejection rates for the four-fermion and \egemm processes, as well as for the signal. A large fraction of background with spectator electrons is removed in this way.

\begin{table}
\centering
\begin{tabular}{| l | c | c |}
\hline
Process & 
\parbox{4cm}{\vspace{.2\baselineskip}Rejection rate by tagging final-state electrons\vspace{.2\baselineskip}} & 
\parbox{4.7cm}{\vspace{.2\baselineskip}Total rejection including \par random Bhabha coincidence\vspace{.2\baselineskip}}  \\ 
\hline
\eemm &  44\percent   &  48\percent   \\
\egemm & 38\percent  & 42\percent  \\
Signal & 0.2\percent  & 7\percent \\
\hline
\end{tabular}
\caption{\label{tab-tagging} Rejection rates through EM shower tagging in the forward region for the most relevant signal and background processes.}
\end{table}

\section{Event selection}
\label{sec-selection}
\subsection{Preselection}
\label{sec-presel}
In order to suppress the influence of the beam-induced background, only reconstructed particles with \pT > 5 GeV were used in the analysis. Under this condition, the preselection of events was made by requiring a reconstruction of two muons in the event, with an invariant mass of the di-muon system in the range 105-145 GeV, as well as the absence of tagged electrons with energy above 200 GeV and polar angle above 1.7\degrees.

\subsection{MVA event selection}
\label{sec-mva}
For the final selection, \ac{MVA} techniques were used based on the \ac{BDT} classifier implemented in the \ac{TMVA} package. The following sensitive observables were used for the classification of events:

\begin{itemize}
   \item Visible energy of the event excluding the energy of the di-muon system, $E_{vis}$,
   \item Transverse momentum of the di-muon system, $\pT(\mumu)$,
   \item Scalar sum of the transverse momenta of the two selected muons, 
         $\pT(\PGm_1)+ \pT(\PGm_2)$,
   \item Boost of the di-muon system, $\beta_{\mumu} = \left | p_{\mumu} \right | / E_{\mumu}$,
   \item Polar angle of the di-muon system, $\theta_{\mumu}$,
   \item Cosine of the helicity angle, 
         $\cos\theta^* = \frac{\vec{p'}(\PGm_1) \vec{p}(\mumu)} 
                {\left| p'(\PGm_1) \right|  \left| p(\mumu) \right|}$,
         where the apostrophe denotes the rest frame of the di-muon system. 
\end{itemize}

Distributions of the sensitive observables for signal and all groups of background processes are given in the Appendix. The distributions of the process \nnmm are very similar to those of the signal, implying that this process is irreducible by kinematic selections. The distributions of the process \ggnnmm show small differences with respect to the signal. All processes with one or two spectator electrons show significant differences from the signal, primarily in the distribution of the visible energy. Further, for these processes the distribution of $\pT(\mumu)$ exhibits a peak at low values of $\pT(\mumu)$. This peak corresponds to events in which the di-muon system recoils against electron spectators or outgoing photons that are emitted below the angular cut of the very-forward EM-shower tagging.  

\begin{figure}
\centering
\includegraphics[width=.62\textwidth]{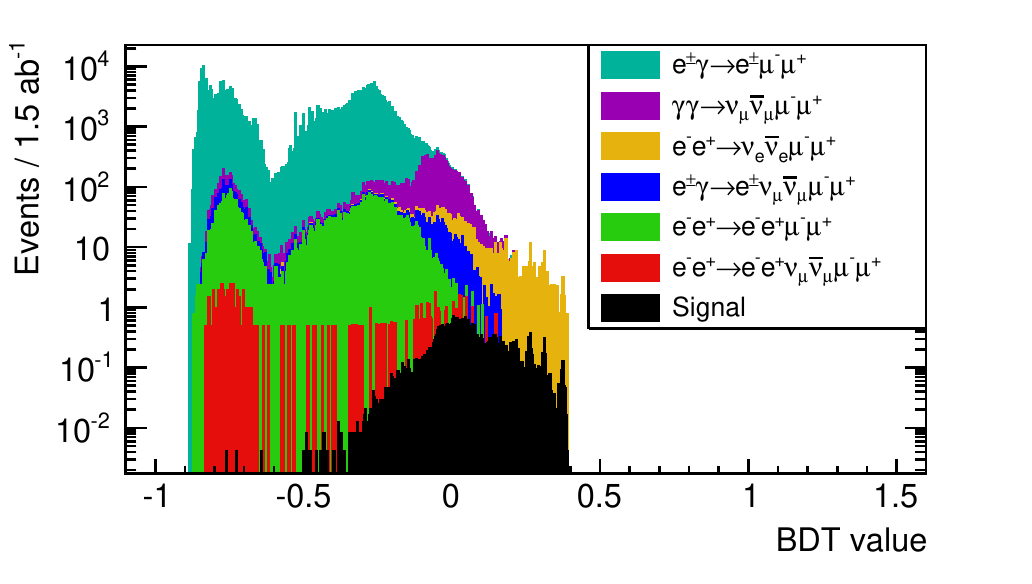}
\includegraphics[width=.37\textwidth]{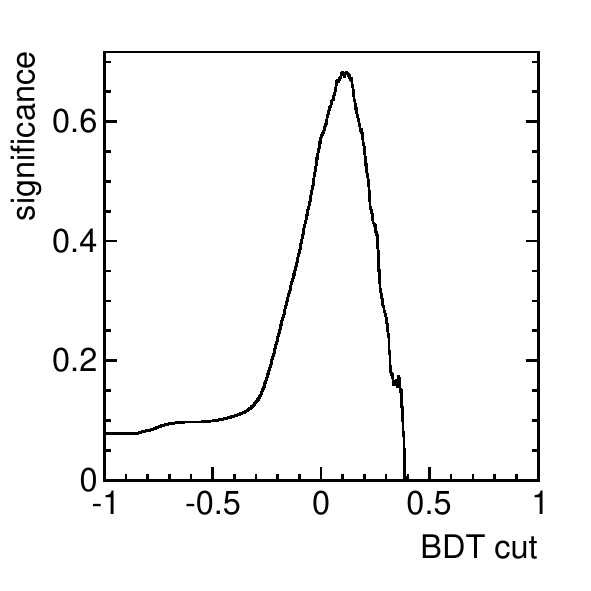}

\hspace{.18\textwidth}(a)\hspace{.45\textwidth}(b)
\caption{\label{fig-BDT} Stacked histograms of the \ac{BDT} classifier distribution for the signal and the background processes (a) and the significance as a function of the \ac{BDT} cut position (b). 
}
\end{figure}

The distribution of the \ac{BDT} classifier variable for the signal and the main background processes is shown in Figure \ref{fig-BDT} (a). Clearly, the largest fraction of background events is very well separated from the signal, with the exception of \ggnnmm and \egennmm events that are separated to some extent, and the irreducible process \nnmm that shows almost the same distribution as the signal. The classifier cut position was selected to maximise the significance, defined as $N_s / \sqrt{N_s + N_b}$, where $N_s$ and $N_b$ are the number of selected signal and background events, respectively. A plot of significance as a function of the position of the \ac{BDT} cut is shown in Figure \ref{fig-BDT} (b). The optimal cut position was found at \ac{BDT} = 0.098. 

\begin{figure}
\centering
\includegraphics[viewport = 0 10 454 170]{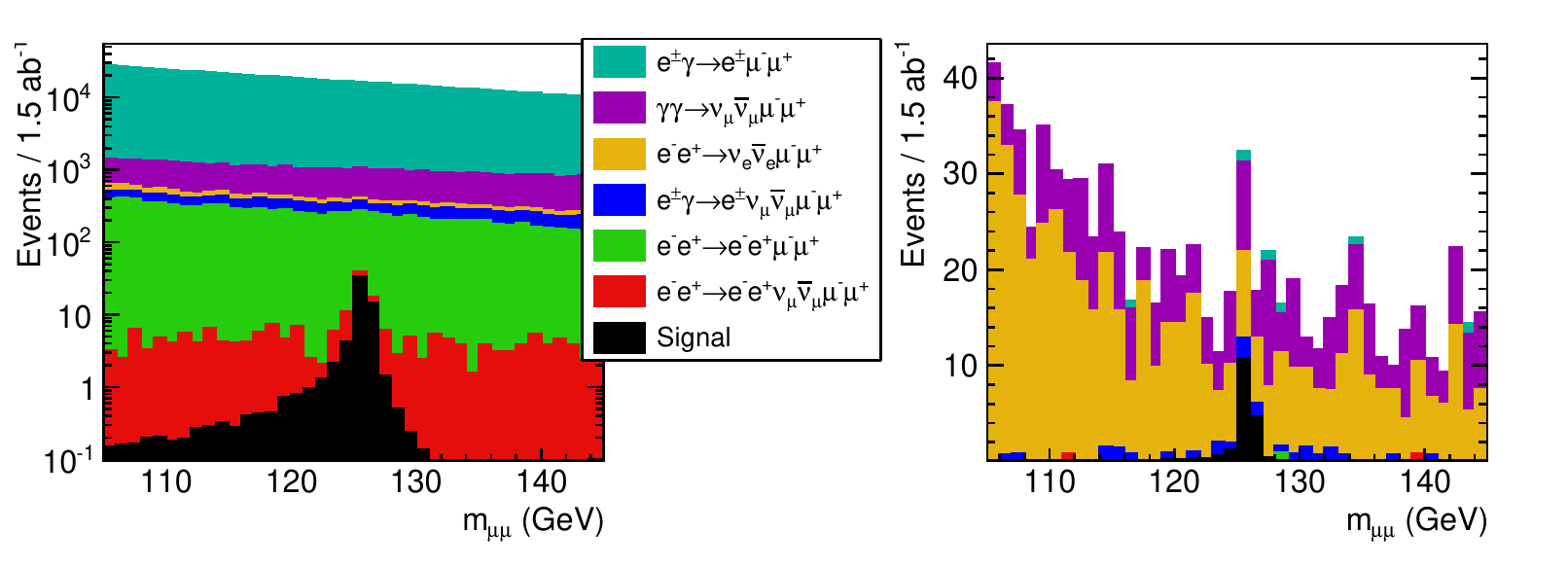}

\hspace{.05\textwidth}(a)\hspace{.55\textwidth}(b)
\caption{\label{fig-stackplot} Stacked histograms of the Di-muon invariant mass distributions with preselecton only (a) and after \ac{BDT} selection (b).}
\end{figure}

Distributions of the di-muon invariant mass are shown in Figure \ref{fig-stackplot}. Figure \ref{fig-stackplot} (a) includes all events that pass the preselection, while Figure \ref{fig-stackplot} (b) shows all events passing the \ac{BDT} selection, as well. All samples were normalised to an integrated luminosity of $1.5\, \abinv$. The \ac{MVA} selection efficiency for the signal is 32\percent. The overall signal efficiency including reconstruction, preselection, losses due to coincident tagging of Bhabha particles and the \ac{MVA} is 26\percent, resulting in an expected number of 20 signal events.

\section{Di-muon invariant mass fit}
\label{sec-fit}
The key observable for the determination of $\sigmaBR$ is the number of selected signal events $N_s$ passing the final selection. 
\begin{equation}
\label{eq-sigma}
	\sigmaBR = \frac{N_s}{L \cdot \epsilon_s}
\end{equation}
where $L$ is the integral luminosity given in units of $\unit{fb}^{-1}$ and $\epsilon_s$ is the total counting efficiency for the signal, including the reconstruction, preselection and \ac{MVA} selection efficiencies.

In the experiment, the number of signal events is determined by fitting the model of the combined \ac{PDF}, $f(m_{\mumu})$, of the di-muon invariant mass, $m_{\mumu}$, for the signal and the background to the measured $m_{\mumu}$ distribution,
\begin{equation}
\label{eq-s+b-model}
	f(m_{\mumu}) = N_s f_s(m_{\mumu}) + N_b f_b(m_{\mumu})
\end{equation}
where $f_s(m_{\mumu})$ and $f_b(m_{\mumu})$ are the \ac{PDF} for the signal and the background, respectively, and $N_s$ and $N_b$ are the respective numbers of signal and background events in the fitting window. The extraction of $f_s(m_{\mumu})$ and $f_b(m_{\mumu})$ from simulated data is described in Section \ref{sec-pdf}.

In order to estimate the statistical uncertainty of the present analysis, 5000 Toy \ac{MC} experiments were performed where pseudo-data were obtained by randomly picking signal $m_{\mumu}$ values from the fully simulated signal sample, while background $m_{\mumu}$ values were randomly generated from the background \ac{PDF}. The sample sizes, $N_s$ and $N_b$, were sampled in each Toy \ac{MC} experiment from the Poisson distribution with the respective mean values $\left< N_{s/b} \right> = L \cdot \sigma_{s/b} \cdot \epsilon_{s/b}$, where the integral luminosity $L$ is $1.5\;\abinv$, $\epsilon_{s/b}$ is the total selection efficiency, and $\sigma_{s/b}$ the corresponding cross-section for the signal and the background, respectively. For each Toy \ac{MC} experiment, the $m_{\mumu}$ distribution is fitted by the function (\ref{eq-s+b-model}), and the RMS of the resulting distribution of $N_s$ over all Toy \ac{MC} experiments is taken as the estimate of the statistical uncertainty of the measurement.

\subsection{Signal and background PDF}
\label{sec-pdf}

The signal and background \ac{PDF} were extracted by fitting to fully simulated datasets after the preselection and the \ac{MVA} selection. 

\begin{figure}[h!]
\centering
\includegraphics[width=8cm,clip, trim = .6cm 6.6cm .6cm 6.6cm]{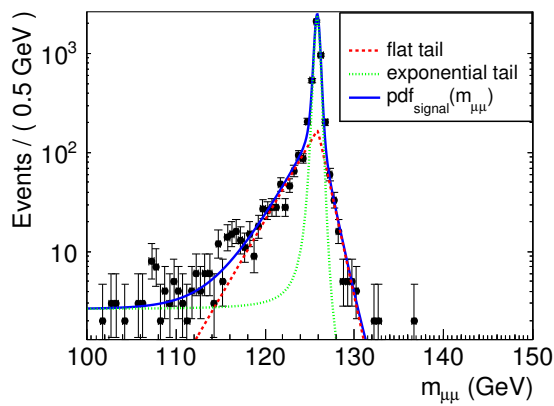}
\caption{\label{fig-SignalFit} Distribution of the invariant mass, $m_{\mumu}$, in the signal sample after \ac{MVA} selection, and the corresponding fit to the data. The fit was performed in the range corresponding to the preselection cut in the analysis. The extension of the line below 105\unit{GeV} is a graphical extrapolation.}
\end{figure}

The signal sample was fitted with an \textit{ad-hoc} function composed of a Gaussian with a flat tail and a Gaussian with an exponential tail (Eq.\ (\ref{eq-pdf-signal}), Figure \ref{fig-SignalFit}). The likelihood fit was performed with unbinned data using \roofit \cite{RooFit:2003}. The results of the fit are listed in Table \ref{tab-sig-pars}.

\begin{align}
f_s &= f_{flat} + C \cdot f_{exp} \nonumber 
\\
f_{flat} &= \left\{ \begin{array}{rl}
		e^{-\frac{ (m_{\mumu} - m_{\PH})^2 }
          {2 \sigma^2 + \beta_L (m_{\mumu} - m_{\PH})^2 } }
	&	m_{\mumu} < m_{\PH} \\
		e^{-\frac{ (m_{\mumu} - m_{\PH})^2 }
          {2 \sigma^2 + \beta_R (m_{\mumu} - m_{\PH})^2 } } 
	& 	m_{\mumu} > m_{\PH} 
	\end{array} \right. \label{eq-pdf-signal} \\
f_{exp} &= \left\{ \begin{array}{rl}
		e^{-\frac{ (m_{\mumu} - m_{\PH})^2 }
          {2 \sigma^2 + \alpha_L |m_{\mumu} - m_{\PH}| } }
	&	m_{\mumu} < m_{\PH} \\
		e^{-\frac{ (m_{\mumu} - m_{\PH})^2 }
          {2 \sigma^2 + \alpha_R |m_{\mumu} - m_{\PH}| } } 
	& 	m_{\mumu} > m_{\PH} 
	\end{array} \right. \nonumber 
\end{align}

\begin{table}[h!]
\centering
	\begin{tabular}{| l | c |}
		\hline
		Parameter	& fitted value \\
		\hline
		$C$			& $0.076 \pm 0.008$ \\
		$\alpha_L$			& $2.49 \pm 0.19 \unit{GeV}$ \\
		$\alpha_R$			& $0.94 \pm 0.06 \unit{GeV}$ \\
		$\beta_L$			& $0.157 \pm 0.004$ \\
		$\beta_R$			& $0.126 \pm 0.006$ \\
		$m_{\PH}$			& $125.847 \pm 0.006 \unit{GeV}$ \\
		$\sigma$			& $0.246 \pm 0.007 \unit{GeV}$ \\
		\hline
	\end{tabular}
\caption{\label{tab-sig-pars}Fitted parameters of the distribution of the invariant mass, $m_{\mumu}$, of the signal.}
\end{table}
 
The $m_{\mumu}$ distribution from the inclusive background data sample after event selection was fitted with a linear combination of a constant and an exponential term, 

\begin{equation}
f_b = p_0(p_1 e^{p_2(m - m_{\PH})} + (1 - p_1))
\label{eq-bkg-pdf}
\end{equation}

\begin{figure}
\centering
\includegraphics{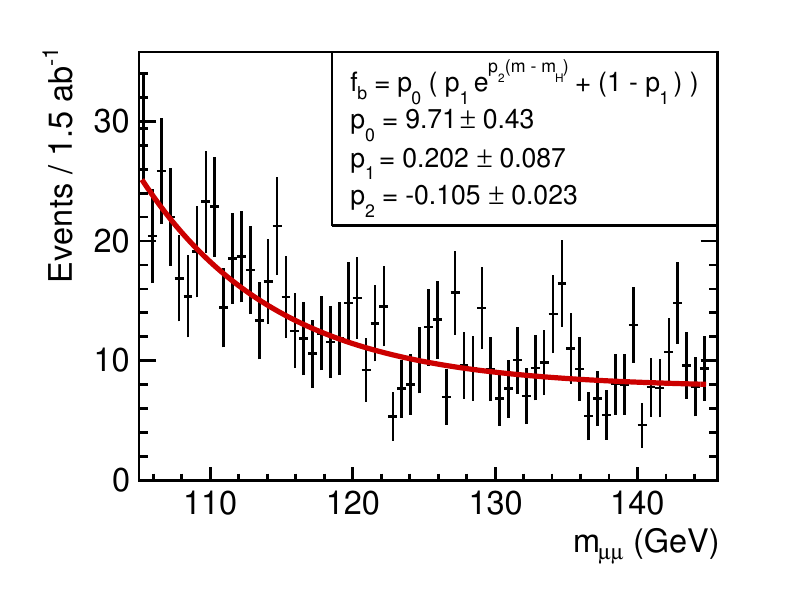}
\caption{\label{fig-bkgFit} Distribution of the invariant mass, $m_{\mumu}$, for the sum of all background processes after event selection. The result of the fit is also shown.}
\end{figure}

The fit results for background are shown in Figure \ref{fig-bkgFit}. As the normalisation to the common integrated luminosity requires different normalisation coefficients for different processes, binned data were used to combine the background processes in a straightforward manner, and binned $\chi^2$ fit was performed. The $\chi^2 / N_{df}$ of the background fit was 62/61.

\begin{figure}
\centering
\includegraphics{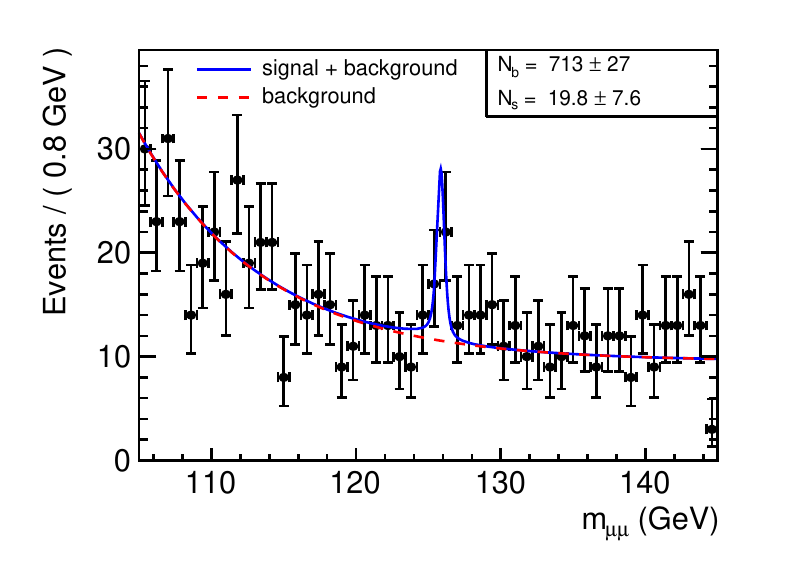}
\caption{\label{fig-TMCfit} Distribution of the invariant mass, $m_{\mumu}$, for the sum of the signal and background samples in one Toy \ac{MC} run, together with the fit of the combined \ac{PDF} model (Eq. (\ref{eq-s+b-model})).}
\end{figure}

The overall function $f(m_{\mumu})$ (\ref{eq-s+b-model}) was fitted to the pseudo data of each Toy \ac{MC} “experiment” using the unbinned likelihood method, and fixing all parameters of $f(m_{\mumu})$ except $N_s$ and $N_b$. An example of a Toy \ac{MC} fit is given in Figure \ref{fig-TMCfit}. 

\subsection{Distribution of the signal count}
\label{sec-distribution}

\begin{figure}
\centering
\includegraphics[width=.49\textwidth]{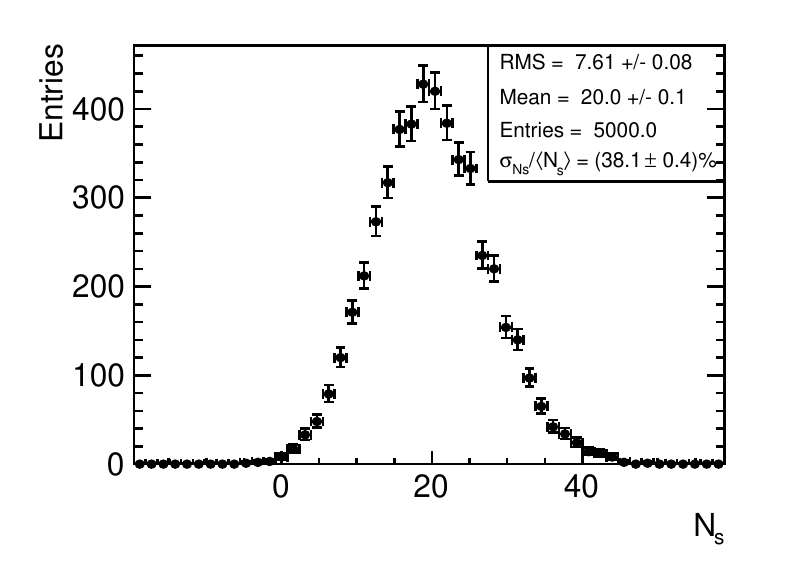}
\includegraphics[width=.49\textwidth]{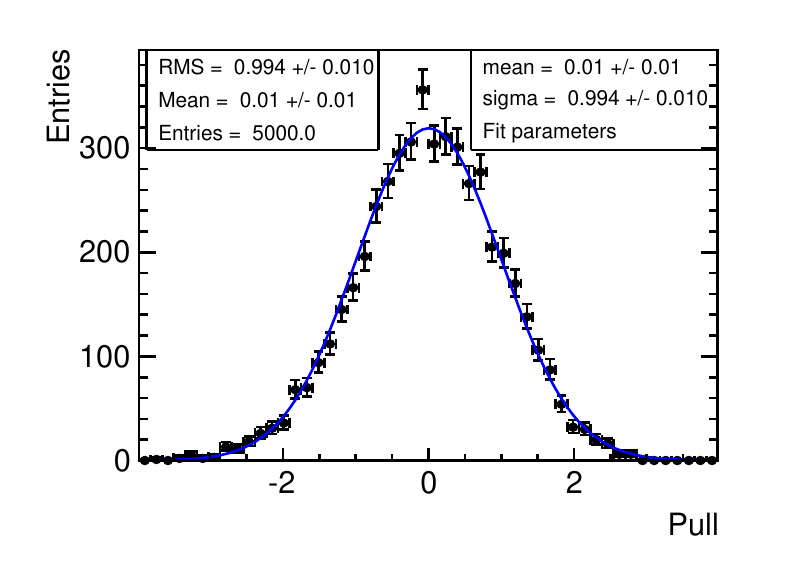}

\hspace{.025\textwidth}(a)\hspace{0.47\textwidth}(b)
\caption{\label{fig-distri} Distribution of the number of signal events in the Toy \ac{MC} experiments (a); the corresponding Pull distribution (b)}
\end{figure}

The RMS deviation of the resulting signal count distribution in 5000 repeated Toy \ac{MC} experiments corresponds to the statistical uncertainty of the measurement (Figure \ref{fig-distri} (a)). The pull distribution (Figure \ref{fig-distri} (b)) is approximately centered at 0 and has an approximate width of 1, confirming that the \ac{PDF}s of the signal and background di-muon invariant mass are adequate.

The relative statistical uncertainty of \sigmaBR is 38\percent. This uncertainty is dominated by the contributions from the limited signal statistics and from the presence of irreducible backgrounds in the \hmumu measurement. With 80\percent left-handed polarisation of the electron beam and 30\percent right handed polarisation for the positrons during the entire operation time at 1.4\unit{TeV}, the Higgs production through \wwfusion would be enhanced by a factor 2.34 \cite{CLIC_snowmass13}. The most important background contribution after the \ac{MVA} selection, the \nnmm process, is enhanced by the same factor because it is also mediated by \PW bosons, which have only left-handed interactions. The remaining background processes are enhanced by a smaller factor, or not enhanced at all. In the conservative estimate, the statistical uncertainty is improved by a factor $1/\sqrt{2.34}$, neglecting the differences in the production ratio for the background processes, as well as the increase in the \ac{MVA} selection efficiency for the signal due to the change in the optimal \ac{BDT} cut. The upper limit of the statistical uncertainty with polarized beams is thus 25\percent. As the polarization in principle affects the distributions of the kinematical observables, a precise estimate is possible only with the full simulation with polarized beams.

To estimate the significance of the signal against the null-hypothesis, another set of 5000 Toy \ac{MC} runs was performed with zero signal count, and $f(m_{\mu\mu})$ (\ref{eq-s+b-model}) was fitted to the pseudo data. The resulting $N_{s}$ distribution was centered on zero with the standard deviation $\sigma_{N_s}(H_0) = 5.4$. Thus, in case the \ac{SM} expected number of events $N_{s,SM} = 20$ is realized in the experiment, the signal significance would be $3.7\,\upsigma$. 

The Higgs coupling to muons, \gHMuMu, is optimally extracted in a global fit procedure taking into account all Higgs measurements both at the 350\unit{GeV} and 1.4\unit{TeV} stages, and extracting all involved Higgs couplings, as well as the total Higgs width in the same procedure. However, even in the suboptimal procedure of extraction of \gHMuMu involving the present measurement alongside measurements giving access to \GH and \gHWW, the dominant contribution to the coupling uncertainty is the statistical uncertainty of the \hmumu measurement. An example of a minimal set of measurements giving model-independent access to \GH and \gHWW is the following: the recoil mass measurement at 350\unit{GeV} giving access to \gHZZ, the $\PH \to \bb$ measurements at both 350\unit{GeV} and 1.4\unit{TeV} stages giving access to the ratio \gHWW/\gHZZ and the $\PH \to \ww$ measurement at 1.4\unit{TeV} giving access to the ratio $\gHWW^4/\GH$ \cite{CLIC_snowmass13}. Assuming that no correlations exist between uncertainties of these measurements, their relative uncertainties add up quadratically to form the relative uncertainty of $\gHMuMu^2$. However, the contributions of measurement uncertainties other than the statistical uncertainty of the \sigmaBR measurement affect the final uncertainty only at the third significant digit, and can be neglected. Thus, the relative uncertainty of \gHMuMu is \couplingunc\percent. A summary of the results of the present analysis is given in Table \ref{tab-results}.

\begin{table}
\centering
\begin{tabular}{ | c | c |}
\hline
$N_s$  & $20.0 \pm 7.6$ \\
$\epsilon_s$  & 26\percent \\
$\frac{\delta(\sigmaBR)}{\sigmaBR}$ & 38\percent \\
$\delta(\gHMuMu)/\gHMuMu$ & \couplingunc\percent \\
\hline
\end{tabular}
\caption{\label{tab-results} Summary of the results of the analysis of the $\sigmaBR$ measurement at 1.4 TeV CLIC with unpolarised beams. All uncertainties are statistical.}
\end{table}

\section{Benefit of a better \pT resolution}
\label{sec-pTres}
\begin{figure}
\centering
\includegraphics{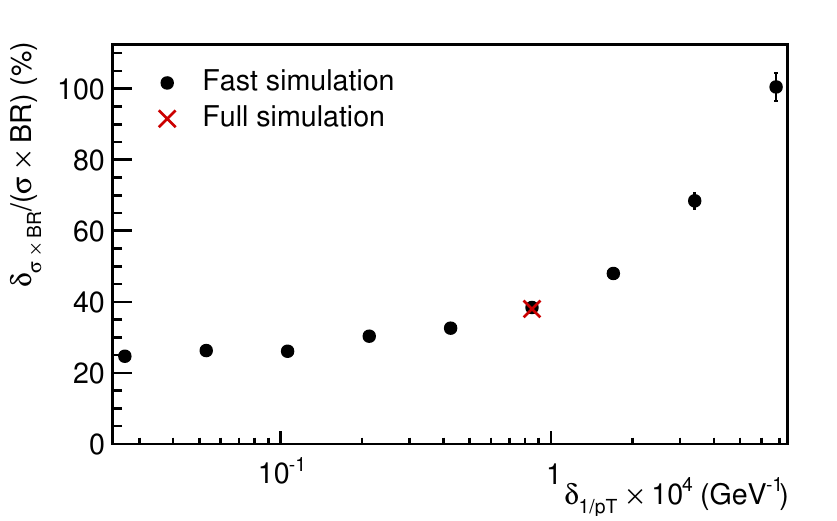}
\caption{\label{fig-pTres} Dependence of the relative statistical uncertainty of the $\sigmaBR$ on the transverse momentum resolution, $\delta_{1/\pT}$, averaged over the signal sample in the whole detector.}
\end{figure}

To estimate the eventual benefit of a better \pT resolution, the analysis was repeated with  substitution of the muon four-momenta reconstructed in the full simulation in the signal sample by the four-momenta obtained by a parametrisation of the momentum resolution ("fast simulation") for several different values of the resolution. Figure \ref{fig-pTres} displays the approximate dependence of the statistical uncertainty of the measurement on the average transverse momentum resolution in the whole detector. It is clear that even a large improvement of the momentum resolution would result in only a moderate improvement of the statistical uncertainty of the measured product of the cross-section and the branching ratio. 

\section{Systematic uncertainties}
\label{sec-systematics}
From Eq.\ (\ref{eq-sigma}) it is clear that uncertainties of the integral luminosity and muon identification efficiency influence the uncertainty of the branching ratio measurement at the systematic level. It has been shown in \cite{Luk13} that at a 3~TeV CLIC, where the impact of the beam-induced processes is the most severe, luminosity above ca. 75\percent of the nominal \ac{CM} energy can be determined at the permille level using low-angle Bhabha scattering. Below 75\percent of the nominal \ac{CM} energy the luminosity spectrum can be measured with a precision of a few percent using wide-angle Bhabha scattering \cite{Poss14}. About 17\percent of all Higgs events occur at a \ac{CM} energy of the initial \epem pair below 75\percent of the nominal \ac{CM} energy. Having in mind the intrinsic statistical limitations of the signal sample, this source of systematic uncertainty can be considered negligible.

\begin{figure}
\centering
\includegraphics[width=.5\textwidth]{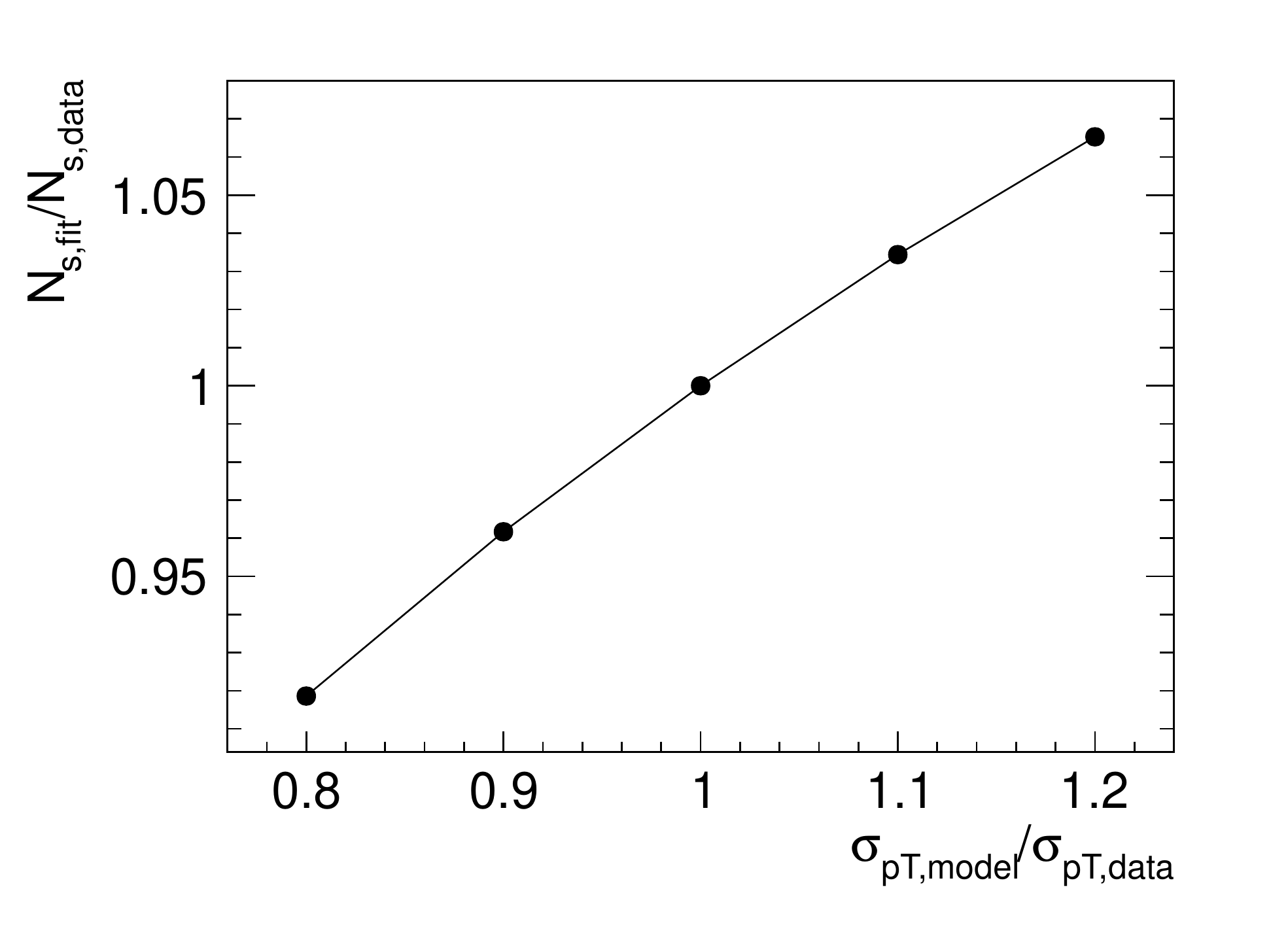}
\caption{\label{fig-bias} Impact of the uncertainty of the muon \pT resolution on signal counting. The relative shift of the signal count is given as a function of the relative shift of the \pT resolution. }
\end{figure}

On the detector side, an important systematic effect is the uncertainty on the transverse momentum resolution, because it directly influences the expected shape of the signal $m_{\mumu}$ distribution. The sensitivity of the signal count to the accuracy of the knowledge of the \pT resolution has been studied by performing the analysis with an artificially introduced error of an exaggerated magnitude on the assumed \pT resolution used to extract the signal $m_{\mumu}$ \ac{PDF}. Results are shown in Figure \ref{fig-bias}. The relative counting bias $\ud N_s / N_s$ per one percent bias of $\sigma_{pT}$ is 0.35\percent.

The uncertainty of the muon identification efficiency will directly influence the signal selection efficiency. In addition, the uncertainty of the muon polar angle resolution impacts the $m_{\mumu}$ reconstruction. Based on the results of the LEP experiments \cite{s12005ema}, it can be assumed that these detector related uncertainties are below a percent.

Because of the forward EM shower tagging, about 7\percent of signal events are rejected by coincidence with the detection of at least one of the final Bhabha particles. This fraction must be precisely calculated taking into account Bhabha event distributions, beam-beam effects, as well as the dependence of the tagging efficiency on energy and angle of incident electrons and photons. This is work in progress \cite{CLICws:Mak14, CLICws:Sailer14, CLICws:Luk14}, but the uncertainty of this effect is also expected to be negligible compared to the statistical uncertainty of this measurement. 

\section{Conclusions}
\label{sec-conclusions}

There is a strong motivation to use precise Higgs measurements at \ac{CLIC} to search for physics beyond the Standard Model. The measurement of Higgs boson couplings are of particular interest. 
The measurement of the branching ratio for the rare \ac{SM}-like Higgs decay into two muons was simulated at a 1.4 CLIC collider with unpolarised beams. The measurement itself tests the muon identification and momentum resolution of the detector. 

It was shown that the measurement of the cross-section times the branching ratio for the Standard Model Higgs decay into two muons can be performed with a statistical uncertainty of 38\percent, assuming 1.5~\abinv integrated luminosity with unpolarized beams. If the same integrated luminosity is collected with 80\percent left-handed polarisation for the electrons and 30\percent right handed polarisation for the positrons, the statistical uncertainty improves to beter than 25\%. The systematic uncertainties are negligible in comparison. The largest contributions to the statistical uncertainty come from the limited statistics of the signal and from the presence of the signal-like backgrounds from \nnmm and \ggnnmm processes. The uncertainty of \sigmaBR of 38\percent translates into the uncertainty of Higgs coupling to muons, \gHMuMu, of \couplingunc\percent.

\section*{Appendix}
\label{sec-app}
\begin{figure}
\centering
\includegraphics[width=\textwidth]{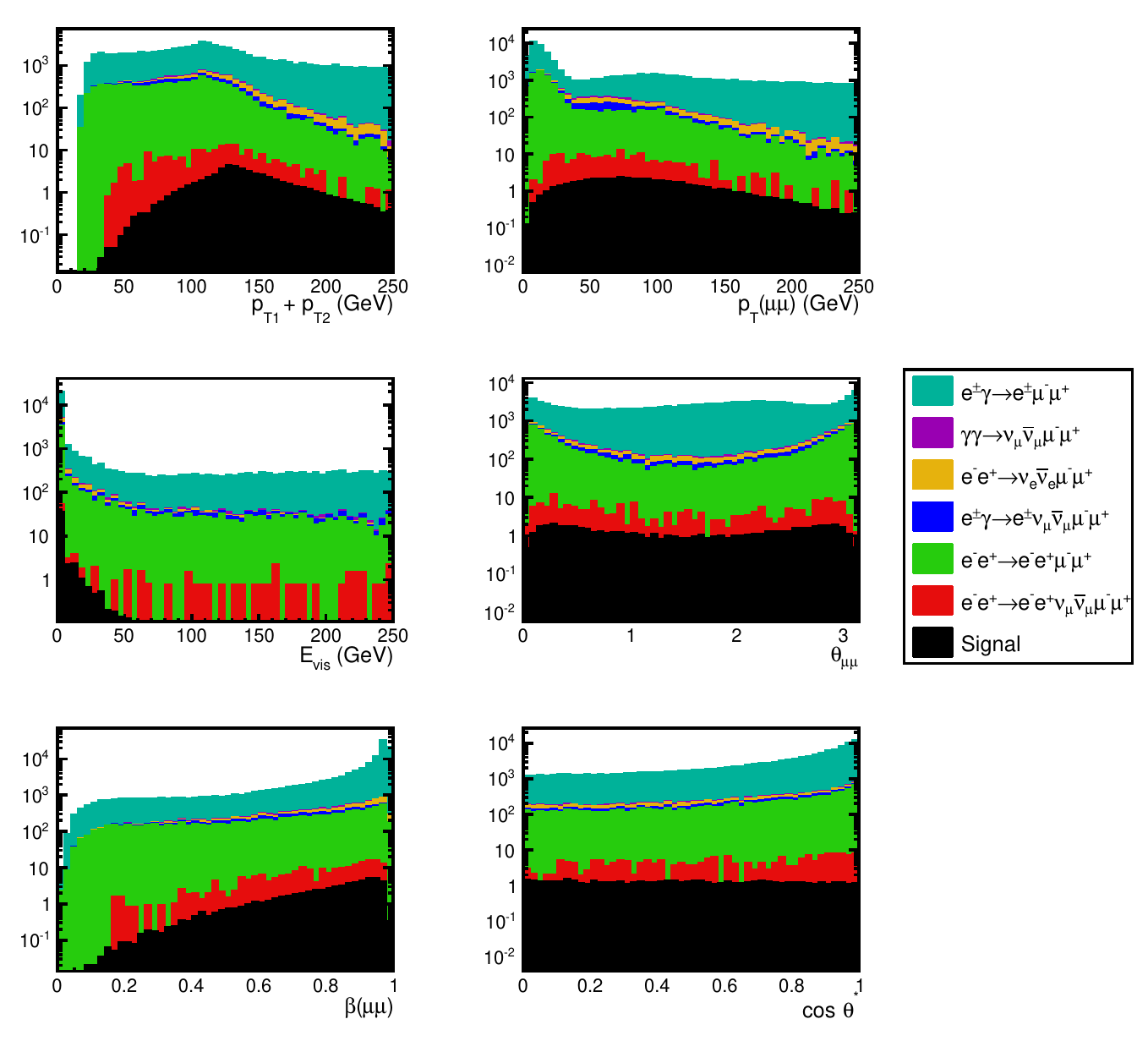}
\caption{\label{fig-variables} Distributions of the sensitive variables used in the \ac{MVA} analysis for events that pass the preselection cuts.}
\end{figure}

\begin{figure}
\centering
\includegraphics[width=\textwidth]{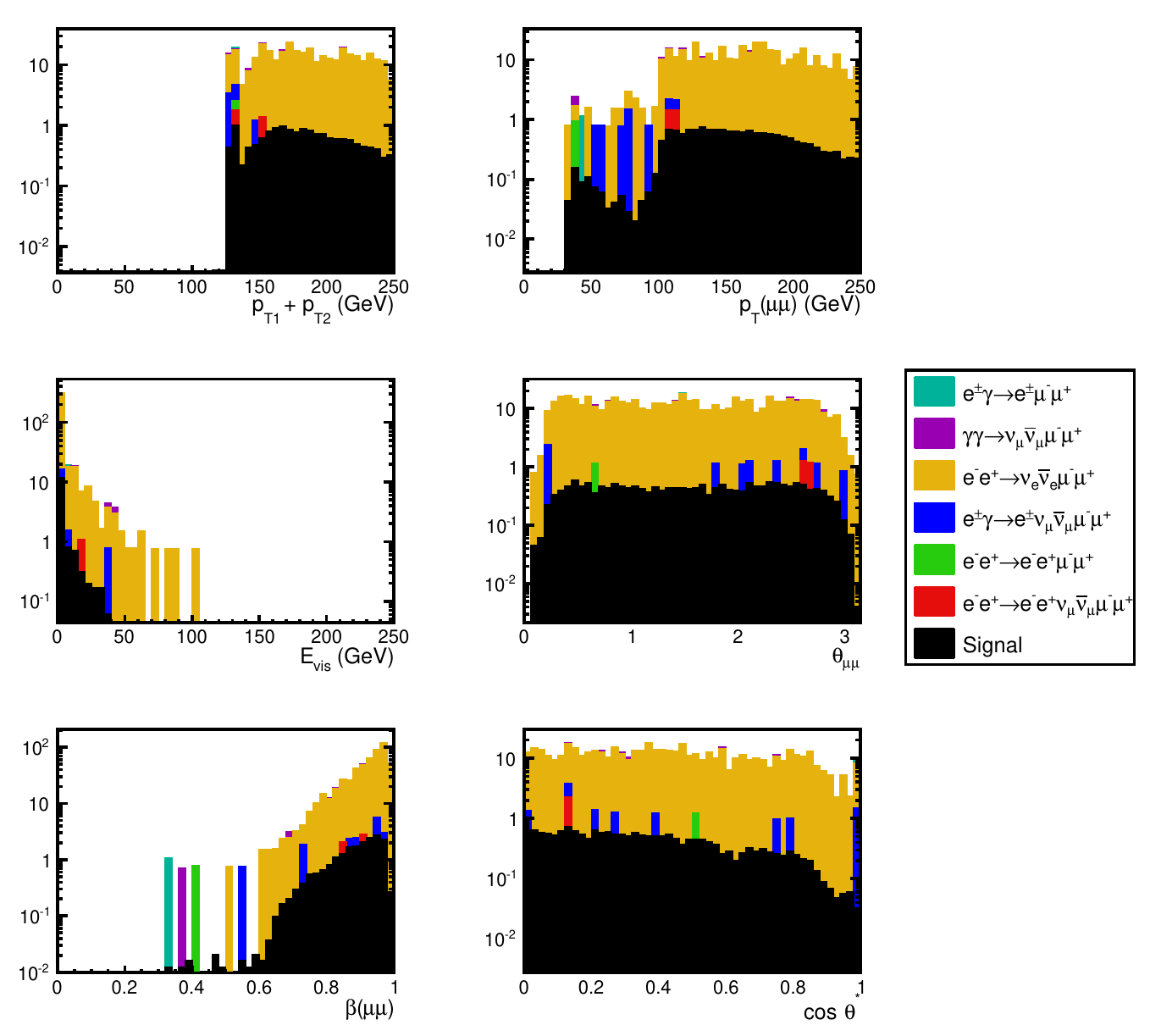}
\caption{\label{fig-variablesBDT} Distributions of the sensitive variables used in the \ac{MVA} analysis for events that pass the preselection and the \ac{MVA} selection.}
\end{figure}

Figures \ref{fig-variables} and \ref{fig-variablesBDT} show the distributions, before and after the \ac{MVA} selection, respectively, of the sensitive variables used in the \ac{MVA} analysis to separate the signal from the background. 

\clearpage
\printbibliography[title=References]

\end{document}